# Optical emission from a kilonova following a gravitational-wave-detected neutron-star merger

Iair Arcavi[1,2], Griffin Hosseinzadeh[1,2], D. Andrew Howell[1,2], Curtis McCully[1,2], Dovi Poznanski[3], Daniel Kasen[4,5], Jennifer Barnes[6], Michael Zaltzman[3], Sergiy Vasylyev[1,2], Dan Maoz[3] and Stefano Valenti[7]

[1]*Department of Physics, University of California, Santa Barbara, CA 93106-9530, USA*

[2]*Las Cumbres Observatory, 6740 Cortona Dr Ste 102, Goleta, CA 93117-5575, USA*

[3]*School of Physics and Astronomy, Tel Aviv University, Tel Aviv 69978, Israel*

[4]*Nuclear Science Division, Lawrence Berkeley National Laboratory, Berkeley, CA 94720-8169, USA*

[5]*Departments of Physics and Astronomy, University of California, Berkeley, CA 94720-7300, USA*

[6]*Columbia Astrophysics Laboratory, Columbia University, New York, NY, 10027, USA*

[7]*Department of Physics, University of California, 1 Shields Ave, Davis, CA 95616-5270, USA*

**The merger of two neutron stars has been predicted to produce an optical-infrared transient (lasting a few days) known as a 'kilonova', powered by the radioactive decay of neutron-rich species synthesized in the merger[1-5]. Evidence that short γ-ray bursts also arise from neutron star mergers has been accumulating[6-8]. In models[2,9] of such mergers a small amount of mass ($10^{-4}$-$10^{-2}$ solar masses) with a low electron fraction is ejected at high velocities (0.1-0.3 times light speed) and/or carried out by winds from an**

**accretion disk formed around the newly merged object**[10,11]. **This mass is expected to undergo rapid neutron capture (r-process) nucleosynthesis, leading to the formation of radioactive elements that release energy as they decay, powering an electromagnetic transient**[1-3,9-14]. **A large uncertainty in the composition of the newly synthesized material leads to various expected colours, durations and luminosities for such transients**[11-14]. **Observational evidence for kilonovae has so far been inconclusive as it was based on cases**[15-19] **of moderate excess emission detected in the afterglows of γ-ray bursts. Here we report optical to near-infrared observations of a transient coincident with the detection of the gravitational-wave signature of a binary neutron-star merger and of a low-luminosity short-duration γ-ray burst**[20]. **Our observations, taken roughly every eight hours over a few days following the gravitational-wave trigger, reveal an initial blue excess, with fast optical fading and reddening. Using numerical models**[21], **we conclude that our data are broadly consistent with a light curve powered by a few hundredths of a solar mass of low-opacity material corresponding to lanthanide-poor (a fraction of $10^{-4.5}$ by mass) ejecta.**

GW170817 was detected[22] by the LIGO[23] and Virgo[24] gravitational-wave detectors on 2017 August 17 12:41:04 (UT used throughout; we adopt this as the time of the merger). Approximately two seconds later, a low-luminosity short-duration γ-ray burst, GRB170817A, was detected[25] by the Gamma-ray Burst

Monitor (GBM) on board the *Fermi* satellite. A few hours later, the gravitational wave signal was robustly identified as the signature of a binary neutron star merger 40±8 Mpc away in a region of the sky coincident with the Fermi localization of the γ-ray burst[26] (Fig. 1).

Shortly after receiving the gravitational-wave localization, we activated our pre-approved program to search for an optical counterpart with the Las Cumbres Observatory (LCO) global network of robotic telescopes[27]. Given the size of the LIGO-Virgo localization region (about 30 square degrees) compared to the field of view of our cameras (about 0.2 square degrees), our search strategy involved targeting specific galaxies[28] (chosen from the GLADE catalog; http://aquarius.elte.hu/glade/) at the reported distance range and location area included in the LIGO-Virgo three-dimensional localization[29] (see Methods).

The fifth galaxy on our prioritized list was NGC 4993, an S0 galaxy 39.5 Mpc away[29]. We observed it with one of the LCO 1-meter telescopes at the Cerro Tololo Inter-American Observatory (CTIO) in Chile on 2017 August 18 00:15:23 and detected a new source at right ascension, $\alpha_{2000}$ = 13h 09m 48.07s and declination, $\delta_{2000}$ = -23°22'53.7", not present in archival images of that galaxy (Fig. 2; see Methods for a timeline of the merger and ensuing immediate follow-up). We are one of a few groups who discovered the same source within 45 minutes of each other (see Methods). It was first announced by the Swope team[31] who named it "SSS17a", but here we use the official IAU designation, AT 2017gfo.

Following the detection of this source, we initiated an intensive follow-up campaign with LCO, obtaining multi-band images of AT 2017gfo for several days, taken from each of our three southern sites (the Siding Spring Observatory in Australia, the South African Astronomical Observatory, and the Cerro Tololo Inter-American Observatory in Chile). AT 2017gfo was visible for less than two hours each night due to the proximity of its position on the sky to the sun, but having a multi-site observatory allowed us to obtain three epochs of observations per 24-hour period, capturing the rapid evolution of the event (Fig. 3).

Our densely sampled light curve reveals that the optical transient peaked approximately 1 day after the merger, followed by rapid fading at a rate of about 2 magnitudes per day in the $g$ band, about 1 magnitude per day in the $r$ band, and about 0.8 magnitudes per day in the $i$ band. The rapid luminosity decline is unlike that of any supernova (Extended Data Fig. 4), but is broadly consistent with theoretical predictions of kilonovae (see, for example, refs 2 and 3). From the temporal and spatial coincidence of this event with both a gravitational-wave signal from a binary neutron-star merger and a short-duration γ-ray burst, we conclude that AT 2017gfo is the kilonova associated with the same merger.

We first compare our observations to analytical models from the literature. The short rise time and luminous bolometric peak of more than $3\times10^{41}$ erg s$^{-1}$ (as indicated by blackbody fits to post-peak multi-color data; see Methods) are consistent with a low-opacity ejected mass according to available analytical models[11,32], but the observed high early temperature is not (see Methods).

With this in mind, we compare the observations to detailed numerical radiation transport models of kilonova light curves and spectra[21]. The model parameters are the total ejecta mass, the characteristic expansion velocity, defined as $(2E/M_{ej})^{1/2}$ (where E is the total kinetic energy imparted on the ejecta mass $M_{ej}$), and the mass fraction of lanthanide species, which are crucial in setting the opacity. This model solves the multi-wavelength radiation transport equation using detailed opacities derived from millions of atomic lines, while self-consistently calculating the temperature and ionization/excitation state of the radioactively heated ejecta (see ref. 21 for more details). This allows us to match the per-band light curves, rather than the bolometric luminosity.

This approach produced a better match to our data, reproducing most of the luminosity evolution (except in *g* band; see below) using an ejecta mass of $(2–2.5)\times10^{-2}\,M_\odot$ (where $M_\odot$ is the solar mass), a characteristic ejecta velocity of 0.3c (where c is the speed of light) and a low lanthanide mass fraction of $X_{lan}$ = $10^{-4.5}$ (Fig. 3), corresponding to an effective opacity of $\kappa\lesssim1$ cm$^2$ g$^{-1}$ (similar parameters also fit our optical spectra presented in ref. 33). This is evidence that the merger produced a component of ejecta composed primarily of light (atomic number $A \lesssim 140$) r-process isotopes. In contrast, the lanthanide mass fraction expected from the production of heavy r-process elements is $X_{lan}$ = $10^{-2}$–$10^{-1}$ (ref. 34), corresponding to κ~10 cm$^2$ g$^{-1}$. A substantial mass of ejecta must have thus experienced significant weak interactions, due to shock heating and/or neutrino interactions, which raised the proton to neutron ratio from the initial value in the neutron star. In such a case, the neutrons available for capture would be exhausted before nucleosynthesis could build up a significant abundance of elements with $A \gtrsim 140$.

The discrepancy in *g* band (and the smaller discrepancy in r-band) may be due to a composition gradient in the ejecta (the model[21] we used assumes a uniform composition). A radial gradient in the lanthanide abundance, in which $X_{lan}$ varies from ~$10^{-6}$ in the outermost layers to ~$10^{-4}$ in the interior layers, can lead to faster reddening of the emission[21], which may fit the data better. Even more lanthanide-rich ejecta ($X_{lan} > 10^{-2}$) could be revealed through emission at later times and redder wavelengths than covered by our data[12-14]. Luminous infrared emission (J~17, H~16, K$_s$~15.5 mag; though some of this emission may be contributed by the host galaxy) is indeed found in observations taken 2.5 and 3.5 days after the merger[35]. It is possible that an additional source of radiation, perhaps related to the ɣ-ray burst engine, contributes to the early blue emission, and could provide an alternative explanation for the *g*- and *r*-band discrepancies. Future modeling efforts will need to explore these options and their effects on the predicted light curves.

The discovery of a kilonova coincident with gravitational waves from a binary neutron star merger and with a short burst of γ-rays provides striking evidence in favor of the main theoretical picture of neutron star mergers. These detections confirm that binary neutron-star mergers produce kilonovae with emission properties broadly in agreement with theoretical predictions. Our early optical to near-infrared light curve shows evidence for a lanthanide-poor component of the mass ejected in the merger, and indications for a blue power source in addition to radioactive decay. The rapid optical evolution explains why transient surveys have so far not detected such events, but the upcoming Large Synoptic Survey Telescope will detect the optical emission of hundreds of kilonovae per year out to distances beyond those accessible to current gravitational-wave detectors (see Methods).

**Acknowledgments** We are indebted to W. Rosing and the LCO staff for making these observations possible, and to the LIGO and Virgo science collaborations. We thank L. Singer, T. Piran and W. Fong for assistance with planning the LCO observing program. We appreciate assistance and guidance from the LIGO-Virgo Collaboration-Electromagnetic follow-up liaisons. We thank B. Tafreshi and G. M. Árnason for helping to secure Internet connections in Iceland while this paper was being reviewed. Support for I.A. and J.B. was provided by the National Aeronautics and Space Administration (NASA) through the Einstein Fellowship Program, grants PF6-170148 and PF7-180162 respectively. G.H., D.A.H. and C.M. are supported by the U.S. National Science Foundation (NSF) grant AST-1313484. D.P. and D.M. acknowledge support by Israel Science Foundation grant No. 541/17. D.K. is supported in part by a Department of Energy (DOE)

Early Career award DE-SC0008067, a DOE Office of Nuclear Physics award DE-SC0017616, and a DOE SciDAC award DE-SC0018297, and by the Director, Office of Energy Research, Office of High Energy and Nuclear Physics, Divisions of Nuclear Physics, of the US Department of Energy under contract number DE-AC02-05CH11231. This research used resources of the National Energy Research Scientific Computing Center, a DOE Office of Science User Facility supported by the Office of Science of the US Department of Energy under contract number DE AC02-05CH11231. This research has made use of the NASA/IPAC Extragalactic Database (NED) which is operated by the Jet Propulsion Laboratory, California Institute of Technology, under contract with NASA. The Digitized Sky Surveys were produced at the Space Telescope Science Institute (STScI) under U.S. Government grant NAG W-2166. The UK Schmidt Telescope was operated by the Royal Observatory Edinburgh, with funding from the UK Science and Engineering Research Council (later the UK Particle Physics and Astronomy Research Council), until 1988 June, and thereafter by the Anglo-Australian Observatory. Supplemental funding for sky-survey work at the STScI is provided by the European Southern Observatory.

**Author Contributions** I.A. is PI of the LCO gravitational wave follow-up program, he initiated and analyzed the observations presented here and wrote the manuscript. G.H. helped with the LCO alert listener and ingestion pipeline, with follow-up observations and image analysis, and performed the blackbody fits. D.A.H. is the LCO-LIGO liaison, head of the LCO supernova group, and helped with the manuscript. C.M. assisted with obtaining and analyzing data, and helped with the LCO alert listener. D.P. helped design the LCO follow-up program, assisted with the galaxy prioritization pipeline and contributed to the manuscript. D.K and J.B. developed theoretical models and interpretations. M.Z. built the galaxy prioritization pipeline. S. Vasylyev built the LCO alert listener and ingestion pipeline. D.M. helped in discussions and with the manuscript. S. Valenti helped with image analysis and with the manuscript.

**Author Information** The authors declare that they have no competing financial interests. Correspondence and requests for materials should be addressed to I.A. (arcavi@gmail.com).

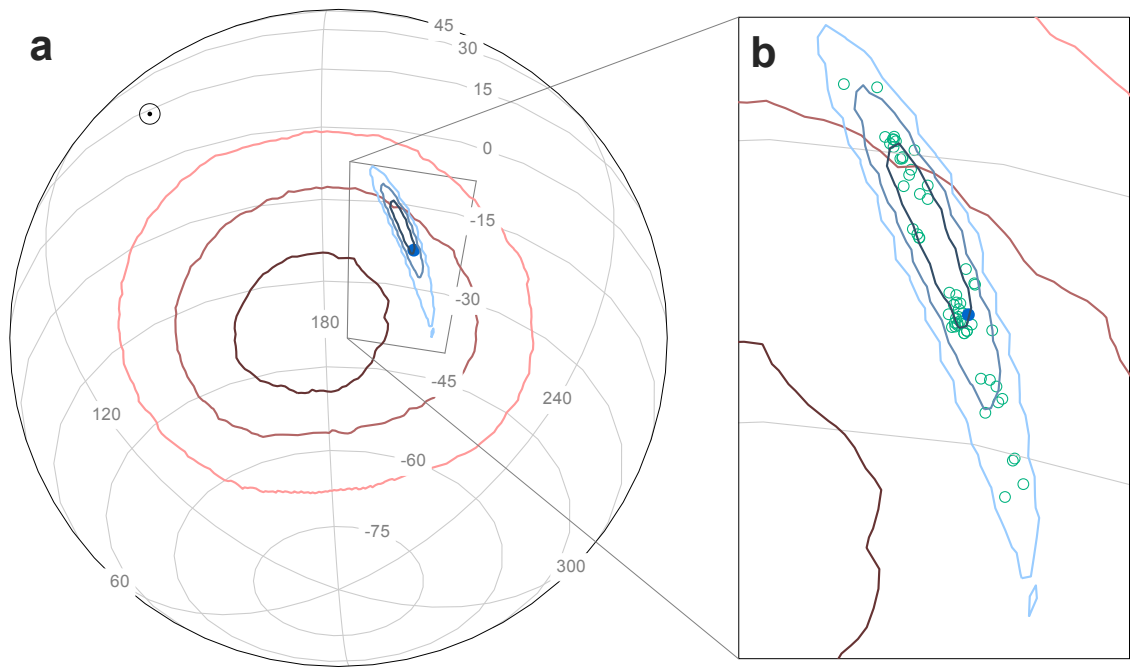

**Figure 1 | Localizations of the gravitational wave, the γ-ray burst and the kilonova on the sky. a**, Our localization of the kilonova AT 2017gfo is shown by the filled circle, together with the localization of GW170817 (blue contours)[26] and that of GRB170817A (red contours)[25]. The contours indicate 1σ, 2σ and 3σ confidence bounds. Representative right ascension and declination values are shown. The position of the sun is indicated by the symbol ☉. **b**, a more detailed view of the kilonova region. Empty circles indicate the locations of other galaxies searched by our LCO follow-up program[36].

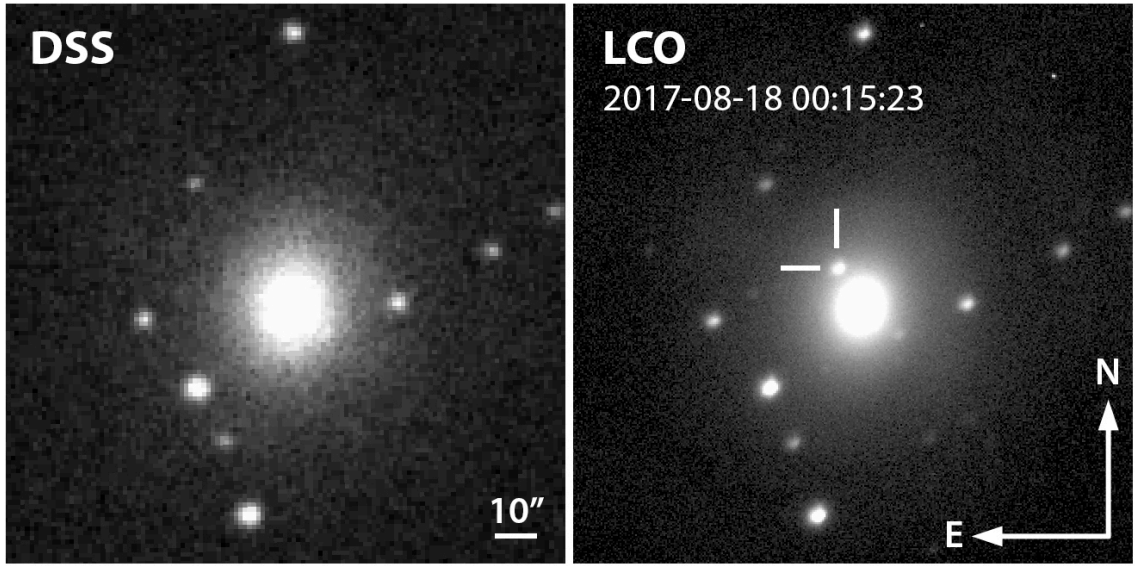

**Figure 2 | LCO discovery image of the kilonova AT 2017gfo in the galaxy NGC 4993.** The *w*-band LCO image (right), centered on NGC 4993, clearly shows a new source (marked with white ticks) compared to an archival image (left) taken on 1992 April 9 with the RG610 filter as part of the AAO-SES survey, retrieved via the Digitized Sky Survey (DSS).

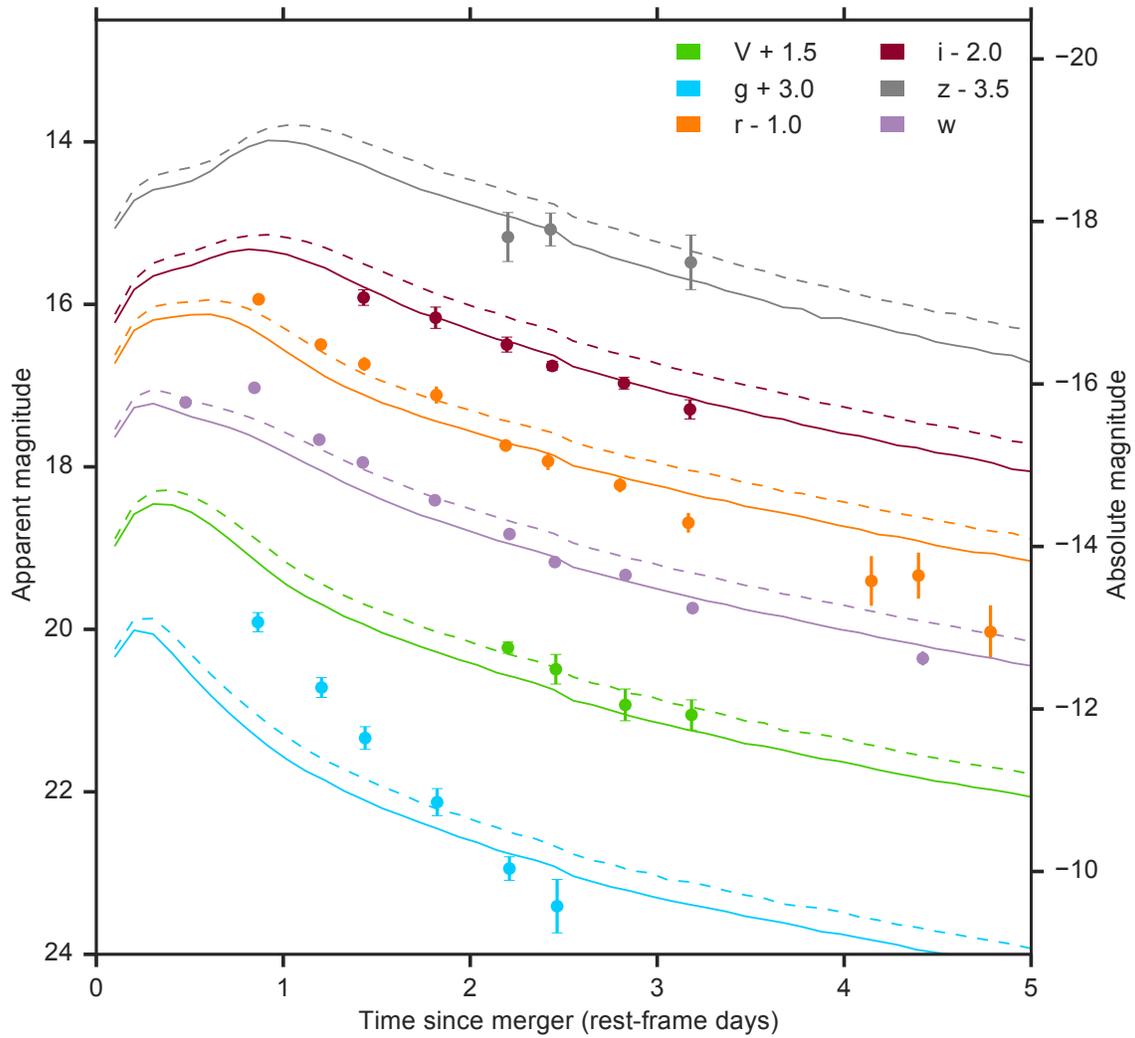

**Figure 3 | LCO light curves of the kilonova AT 2017gfo.** Our rapid-response high-cadence follow-up constrains the peak of the light curve to approximately 1 day after the merger. Numerical radioactive-decay-powered kilonova models[21] are shown for an ejecta mass of $2\times10^{-2}\,M_\odot$ (solid lines) and $2.5\times10^{-2}\,M_\odot$ (dashed lines), a characteristic ejecta velocity of $0.3c$ and a low lanthanide fraction of $10^{-4.5}$. Error bars denote $1\sigma$ uncertainties. Data from the same site, filter and night are binned for clarity. Magnitudes are corrected for host-galaxy contamination using image subtraction, and for Milky-Way extinction.

**Methods**

**Gravitational wave follow-up strategy and kilonova discovery.** Las Cumbres Observatory (LCO)[27] consists of 20 telescopes (two 2-meter, nine 1-meter and nine 0.4-meter in diameter) at six sites around the world, operated robotically as one network using dynamical scheduling software. As stated in the main text, we use a galaxy-targeted follow-up strategy rather than a tiling one[28]. Our galaxy selection strategy prioritizes galaxies that are at higher probability locations and distances in the gravitational wave localization region[26], that have a higher intrinsic *B*-band luminosity (indicative of higher mass), and in which LCO is more likely to be sensitive to a kilonova. More details are provided in ref. 36. The timeline of the discovery, immediate follow-up and the visibility of NGC 4993 are depicted in Extended Data Figure 1. In addition to our detection, AT 2017gfo was independently detected by the Swope, DECam, DLT40, MASTER and VISTA groups[31,37-43] (see also Lipunov V. *et al.*, manuscript in preparation).

**Photometry.** Images from the LCO 1-meter telescopes were pre-processed using the Python-based *BANZAI* pipeline. Photometry was then extracted using the PyRAF-based *LCOGTsnpipe* pipeline[44] by performing image subtraction[45] followed by PSF fitting. We use images taken after the kilonova faded below our detection limits as subtraction references. Our *V*-band data are calibrated to the AAVSO Photometric All-Sky Survey[46] in the Vega system, *grizw*-band data are calibrated to the AB system using SDSS fields observed on the same night as AT 2017gfo, with *w* band (which is a broad *g*+*r*+i band) treated as *r* band. We correct all photometry for Milky Way extinction[47] retrieved via the NASA/IPAC Extragalactic Database (NED; http://ned.ipac.caltech.edu). We adopt a Tully-Fisher distance of 39.5 Mpc (distance modulus of 32.98 mag)[29] to NGC 4993 retrieved via NED.

**Blackbody Fits.** Kilonovae are expected to display roughly blackbody emission (perhaps with a steeper fall-off at short wavelengths due to line blanketing)[10,12,13,33]. We fitted a blackbody spectrum to each epoch containing data in more than two bands (excluding w band) using Markov Chain Monte Carlo (MCMC) simulations through the Python *emcee* package[48] (Extended Data Fig. 2). We find that the photospheric radius remains roughly constant during the first few days after peak at a value of about $5\times10^{14}$ cm while the temperature declines from about 6,500 K 1.4 days after peak to about 4,000 K 2.5 days after peak (Extended Data Fig. 3). We calculate the bolometric luminosity of the blackbody and take that to be the bolometric luminosity of the event.

**Comparison to Supernova Light Curves.** AT 2017gfo peaks at an absolute magnitude fainter than most supernovae, but comparable to that of some type IIb supernovae, and to plateau luminosities of type IIP supernovae (e.g. ref 49). However, AT 2017gfo evolves faster than any known supernova. In Extended Data Figure 4 we compare it to standard type Ia and type Ib/c light curves[50,51], as well as to some of the most rapidly evolving supernovae known[52,53], SN 2002bj and SN 2010X. We also plot the plateau drop phase of the prototypical type IIP supernova[54] SN 1999em. Type IIP supernova light curves have a ~100 day plateau, followed by a rapid drop in luminosity as the power source changes from shock heating to radioactive decay of $^{56}$Co. Still, this sharp decline is slower than the decline in AT 2017gfo. In Extended Data Figure 4 we also plot the DLT40 non-detection pre-discovery limits[55,56] of AT 2017gfo, which further rule out a type IIP supernova origin.

**Fits to Analytical Kilonova Models.** The basic predictions for the peak time, luminosity and temperature of a kilonova, assuming a spherically symmetric, uniform mass distribution for an ejecta in homologous expansion, are[11]:

$$t_{peak} \sim 4.9d \times \left(\frac{M_{ej,-2}\kappa_{10}}{v_{ej,-1}}\right)^{1/2} \quad (1)$$

$$L_{peak} \sim 2.5 \times 10^{40} erg\ s^{-1} \times M_{ej,-2} \left(\frac{M_{ej,-2}\kappa_{10}}{v_{ej,-1}}\right)^{-\alpha/2} \quad (2)$$

$$T_{peak} \sim 2,200K \times M_{ej,-2}^{-\alpha/8} \kappa_{10}^{-(\alpha+2)/8} v_{ej,-1}^{(\alpha-2)/8} \quad (3)$$

where $M_{ej,-2}$ is the ejecta mass in units of $10^{-2}\ M_\odot$, $\kappa_{10}$ is the opacity of the ejecta mass in units of 10 cm$^2$ g$^{-1}$, $v_{ej,-1}$ is the ejecta velocity in units of 0.1c, and α is the power-law index that describes the time dependence of the energy emitted by radioactive decay. Here we use α=1.3, which is typically assumed for r-process decay[57]. The peak luminosity (Eq. 2) is approximately 1,000 times brighter than a nova, giving kilonovae their name[3] (though some use the more general name 'macronovae')[58].

These simple relations can reproduce the short rise time and bright peak luminosity deduced from the blackbody fits (Extended Data Fig. 5) with an ejecta mass $M_{ej}$ of a few times $10^{-2}$ $M_\odot$ and a low ($\kappa \lesssim 1$ cm$^2$ g$^{-1}$) opacity. However, using these values does not reproduce the observed colors, as it under-predicts the observed temperature (Eq. 3). We use these parameters as starting points for MCMC simulations to fit more sophisticated analytical models[31] based on approximations to numerical relativity simulations. We fitted the models to the bolometric light curve rather than using the model bolometric corrections to fit the per-band light curves, since the corrections are only valid for times $> 2d \times (10^{-2} M_\odot/M_{ej})^{-1/3.2}$ after merger, which would miss much of our data. We fix the heating rate coefficient $\dot{\epsilon}_0 = 1.58 \times 10^{10}$ $erg$ $g^{-1}$ $s^{-1}$ and leave the ejecta mass ($M_{ej}$), the minimum and maximum ejecta velocities ($v_{ej,min}$ and $v_{ej,max}$), the opacity ($\kappa$), and the geometrical parameters ($\theta_{ej}$ and $\Phi_{ej}$) as free parameters. We use the public code provided by ref. 59 for these models and adopt the time-varying thermalization efficiency found by ref. 60. Our MCMC fits converge on an ejecta mass of $(4.02 \pm 0.05) \times 10^{-2}$ $M_\odot$ (1$\sigma$ uncertainties), but do not constrain the ejecta velocities (Extended Data Fig. 6) or the geometrical parameters (in ref. 59 it is demonstrated that in general the geometrical parameters can not be constrained in this model). We compare the individual band magnitudes from this fit, using the bolometric corrections supplied by the model and find that they are redder than the observations. We conclude that even the more sophisticated analytical models[31] (under the stated assumptions for α, $\dot{\epsilon}_0$ and the thermalization efficiency) cannot reproduce the color evolution of our event. As stated for our numerical models[21] in the main text, a composition (and hence opacity) gradient or an additional power source, could explain the color-evolution discrepancy.

**Rates.** Given the light curve properties reported in the main text, we can explore how many AT 2017gfo-like events are expected to be seen by different optical transient surveys, without relying on a gravitational- wave trigger. The number of kilonovae per year N potentially seen in E epochs by a survey covering a fraction f of the sky down to limiting magnitude L and with cadence C days is:

$$N = f \times R \times 10^{0.6(L - \Delta m \times C \times (E-1) - m_p)} \quad (4)$$

where R is the rate of kilonovae per year on the entire sky out to a distance d, $\Delta m$ is the decline rate of the kilonova in magnitudes per day, and $m_p$ is the apparent peak magnitude of the kilonova at distance d (we ignore time dilation effects from an expanding Universe). Using the values from our *r* band data ($m_p$=17, $\Delta m$=1) and assuming R=1, we plot the number of detectable kilonovae in Extended Data Figure 7. We find, for example, that a survey with a limiting magnitude of 21 and sky coverage of 4,000 square degrees with 3-day cadence (similar to the Palomar Transient Factory[61,62]) would have a two-epoch detection of only one kilonova roughly every 2-3 years. The Large Synoptic Survey Telescope (LSST), reaching a magnitude of 24 on roughly half of the sky with 3-day cadence, could obtain three epochs for one kilonova per year, and two epochs for each of 100 kilonovae per year. Equation (4) demonstrates that increasing the cadence of a survey has a larger effect on kilonova detections than increasing the sky coverage. It is therefore likely that LSST could discover even more kilonovae in its "deep drilling" fields.

**Data availability**

The photometric data that support the findings of this study are available in the Open Kilonova Catalog[63], https://kilonova.space. Source data for Figure 3 are provided with the paper.

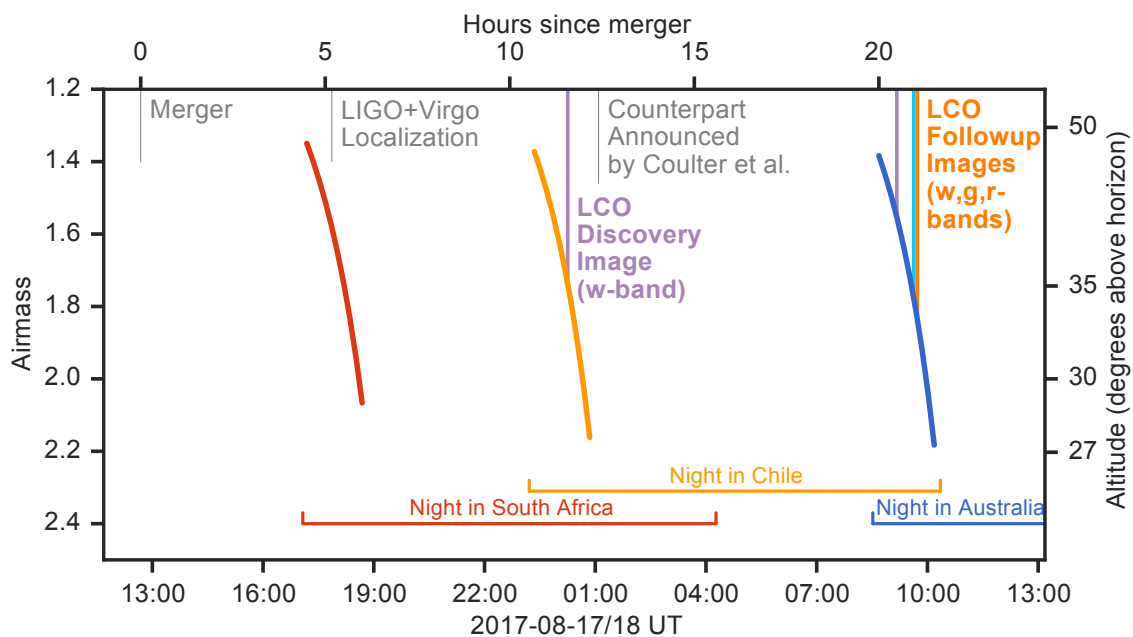

**Extended Data Figure 1 | Timeline of the discovery and the observability of AT 2017gfo in the first 24 hours following the merger.** The curved lines denote the airmass and altitude (in degrees above the horizon) of the position of AT 2017gfo on the sky at each LCO southern site from the start of the night until the hour-angle limit of the LCO 1-meter telescopes. The vertical thick lines denote the times when LCO images were obtained (colors correspond to the different filters as denoted in the legend of Figure 3). AT 2017gfo was observable for approximately 1.5 hours at the beginning of the night. Having three southern sites allowed us to detect the kilonova approximately 6.5 hours after the LIGO-Virgo localization, follow it approximately 10 hours later, and continue to observe it three times per 24-hour period for the following days (Fig. 3). Counterpart announcement is from ref. 31.

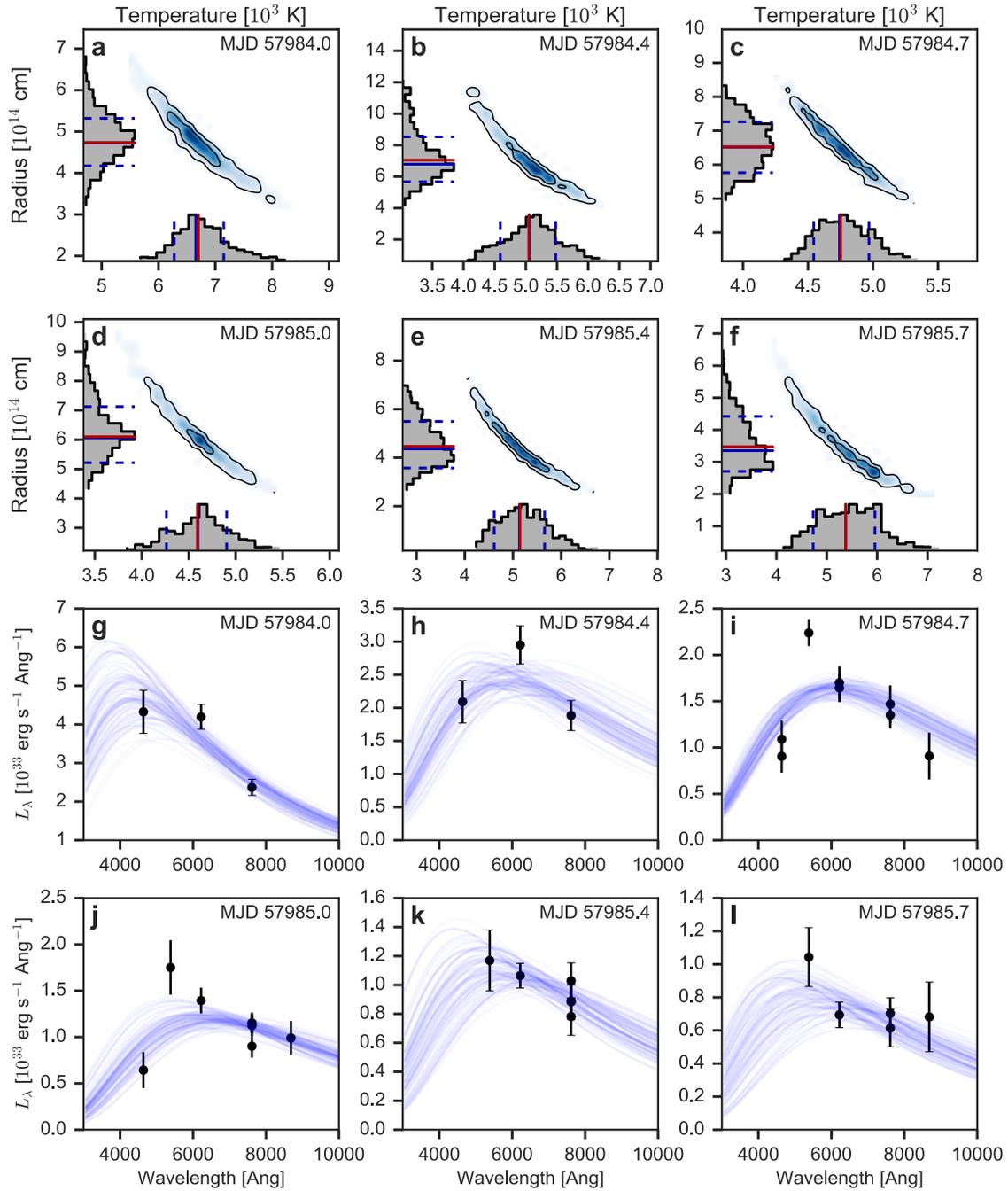

**Extended Data Figure 2 | Blackbody fits.** MCMC parameter distributions (**a**-**f**) and spectral energy distributions (luminosity density as a function of wavelength) with the blackbody fits (**g**-**l**) are shown for the six epochs (noted by their modified Julian dates, MJD) with observations in more than two bands after excluding *w*-

band data. In the parameter distributions, contour lines denote 50% and 90% bounds, the red and blue solid lines overplotted on each histogram denote the mean and median of each parameter distribution (respectively), and the dashed lines denote 68% confidence bounds. Error bars on the luminosity densities denote 1σ uncertainties.

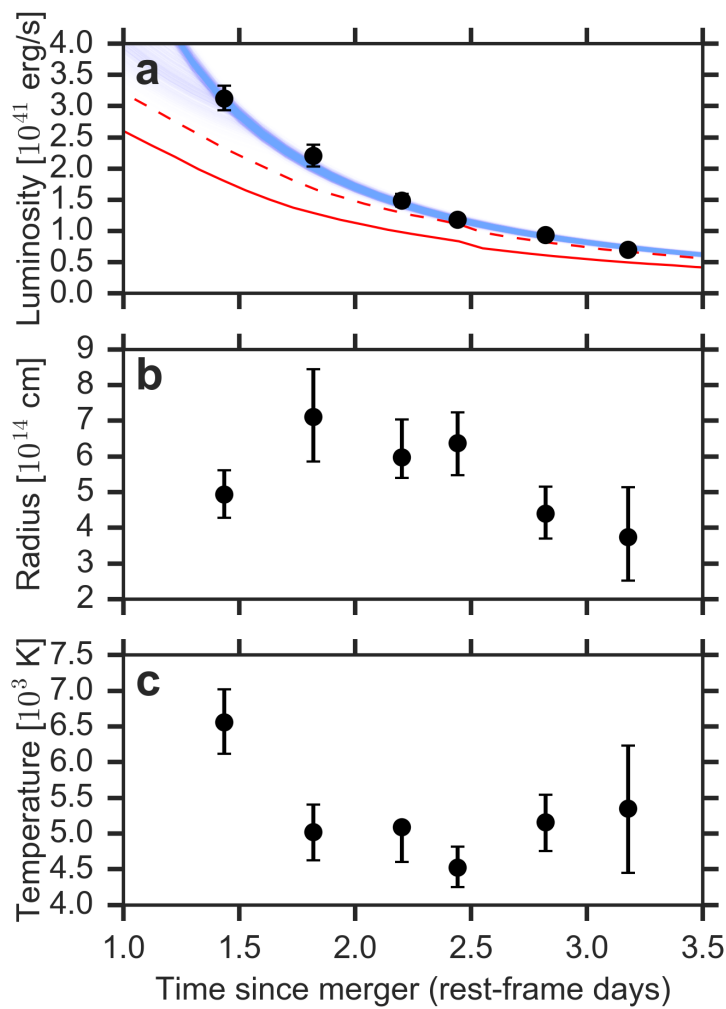

**Extended Data Figure 3 | Bolometric luminosity, photospheric radius and temperature deduced from blackbody fits.** Error bars denote 1σ uncertainties (n=200). The large uncertainties in the later epochs might be due to a blackbody that peaks redward of our available data, thus these data points should be considered to be temperature upper limits. Our MCMC fits of an analytical model[31] to the bolometric luminosity are shown in blue, and the numerical models[21] from Figure 3 are shown in red in the top panel. The numerical models were tailored to fit *Vriw* bands, but not the *g* band, which is driving the high bolometric luminosity at early times.

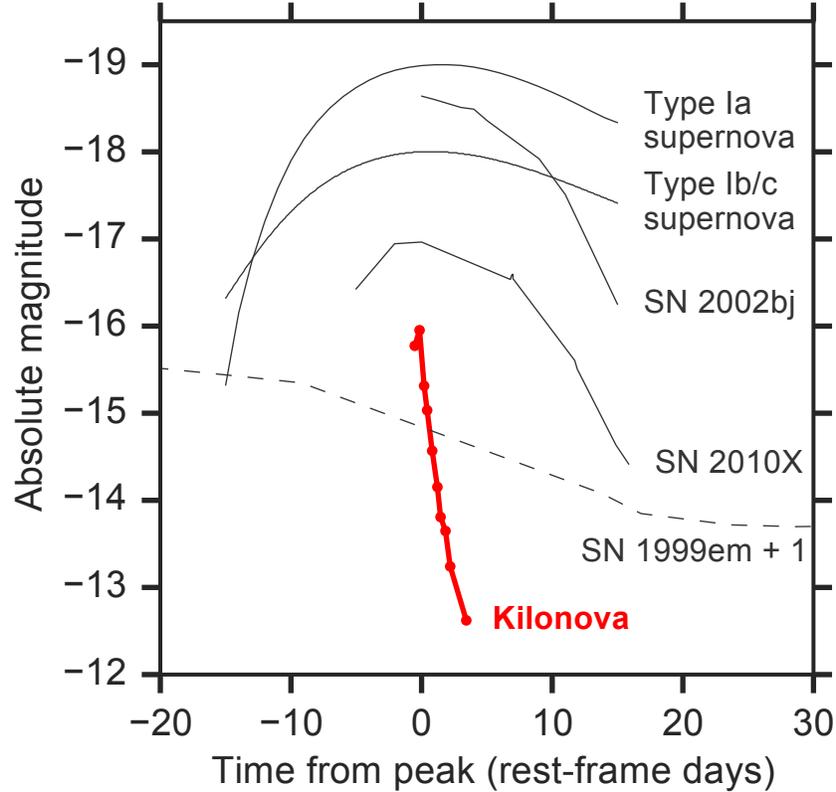

**Extended Data Figure 4 | AT 2017gfo evolves faster than any known supernova, contributing to its classification as a kilonova.** We compare our *w*-band data of AT 2017gfo (red; arrows denote 5σ non-detection upper limits reported by others[55,56]) to *r*-band templates of common supernova types (types Ia and Ib/c normalized to peaks of −19 and −18 mag respectively)[50,51], to *r*-band data of two rapidly-evolving supernovae[52,53] (SN 2002bj and SN 2010X) and to *R*-band data of the drop from the plateau of the prototypical type IIP supernova[54] SN 1999em (dashed line; shifted by one magnitude for clarity).

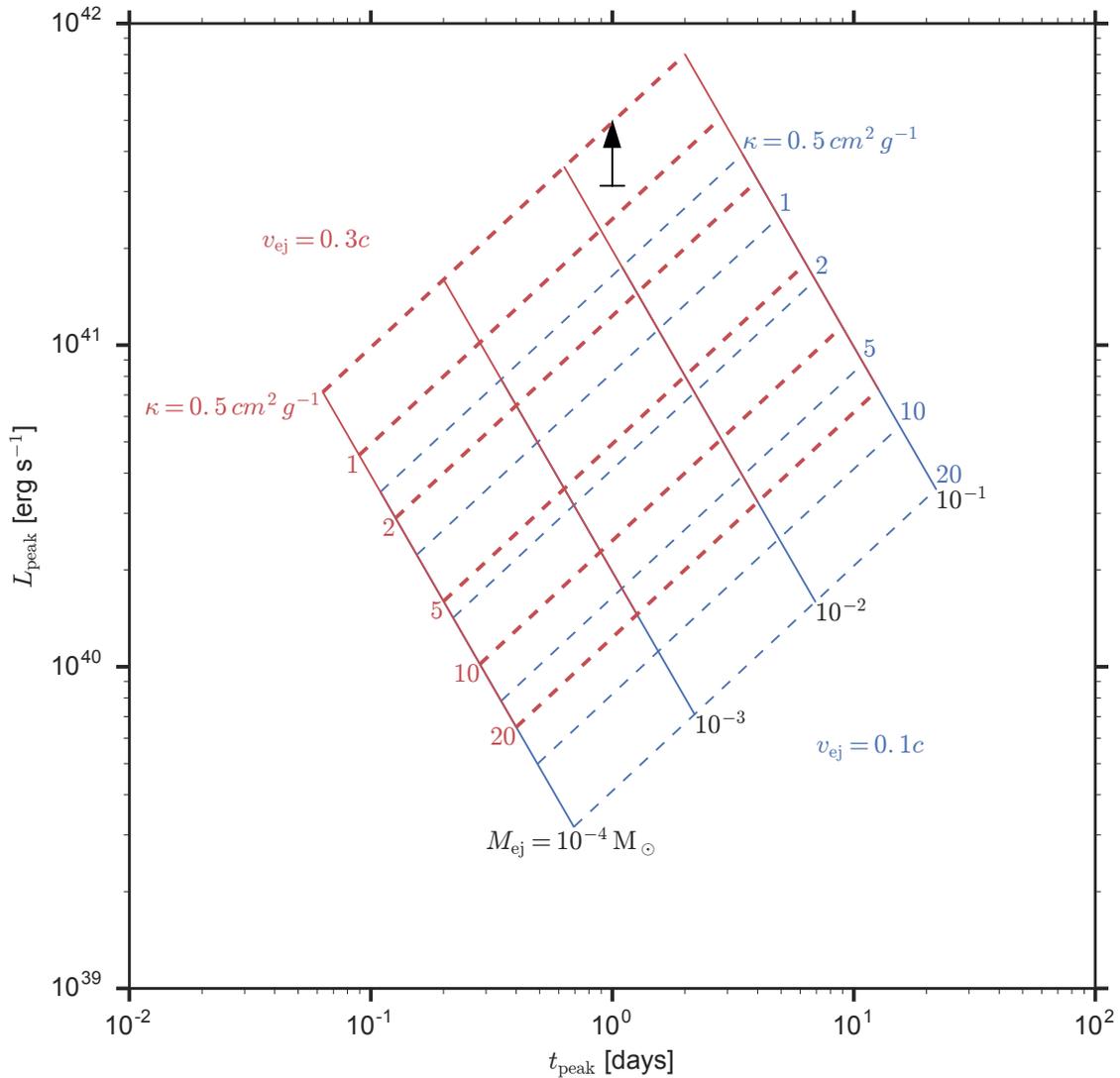

**Extended Data Figure 5 | Peak luminosity and time of AT 2017gfo compared to simple analytical predictions.** The parameters[11] from Equations (1) and (2) are shown for different values of the ejecta mass $M_{ej}$ (solid lines), the opacity $\kappa$ (dashed lines), and for two different ejecta velocities $v_{ej}$ (red and blue). The rise time and peak luminosity of AT 2017gfo (black arrow) can be reproduced by an ejecta velocity $v_{ej} \sim 0.3c$ and a low opacity $\kappa \lesssim 1$ cm$^2$ g$^{-1}$. Matching the data with higher opacities would require higher ejecta velocities.

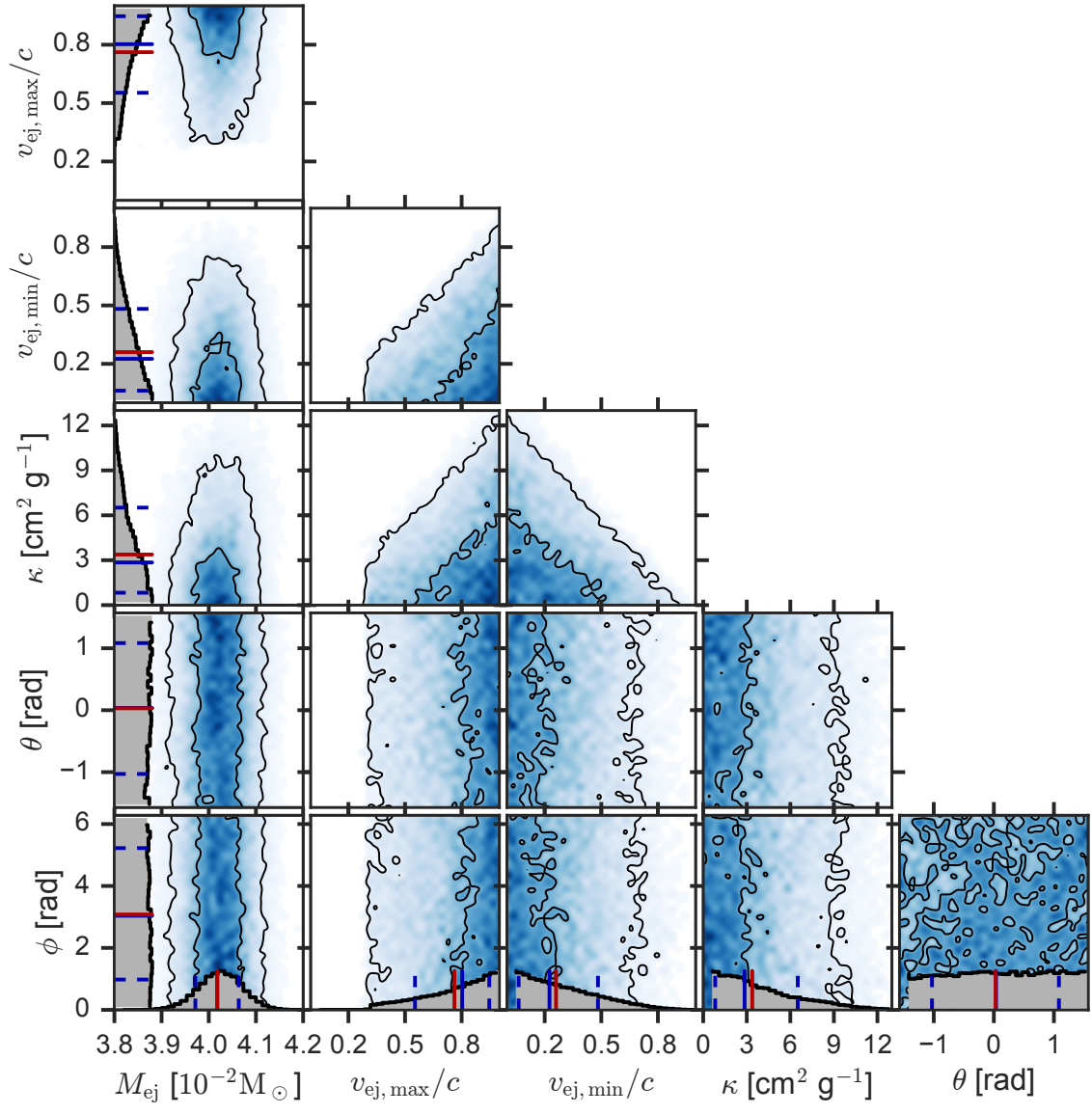

**Extended Data Figure 6 | Parameter distribution for MCMC fits of analytical kilonova models[31] to our bolometric light curve.** The contour lines denote 50% and 90% bounds. The red and blue solid lines overplotted on each histogram denote the mean and median of each parameter distribution (respectively). The dashed lines denote 68% confidence bounds. The fits converge on an ejecta mass of $(4.02 \pm 0.05) \times 10^{-2}$ $M_\odot$ but they do not constrain the velocity (converging on the largest possible range) or the geometrical

parameters ($\theta_{ej}$ and $\Phi_{ej}$), nor do they reproduce the colour evolution of our event (not shown). This indicates that these models may not be entirely valid for AT 2017gfo (although in ref. 59 it is shown that the geometrical parameters can not be constrained either way). Our numerical models[21], on the other hand, which include detailed radiation transport calculations, do provide a good fit to the data (Fig. 3) with $M_{ej} = (2 - 2.5) \times 10^{-2}\,M_\odot$, $v_{ej} = 0.3c$, and a lanthanide mass fraction of $X_{lan}=10^{-4.5}$, corresponding to an effective opacity of $\kappa \lesssim 1\ cm^2\ g^{-1}$.

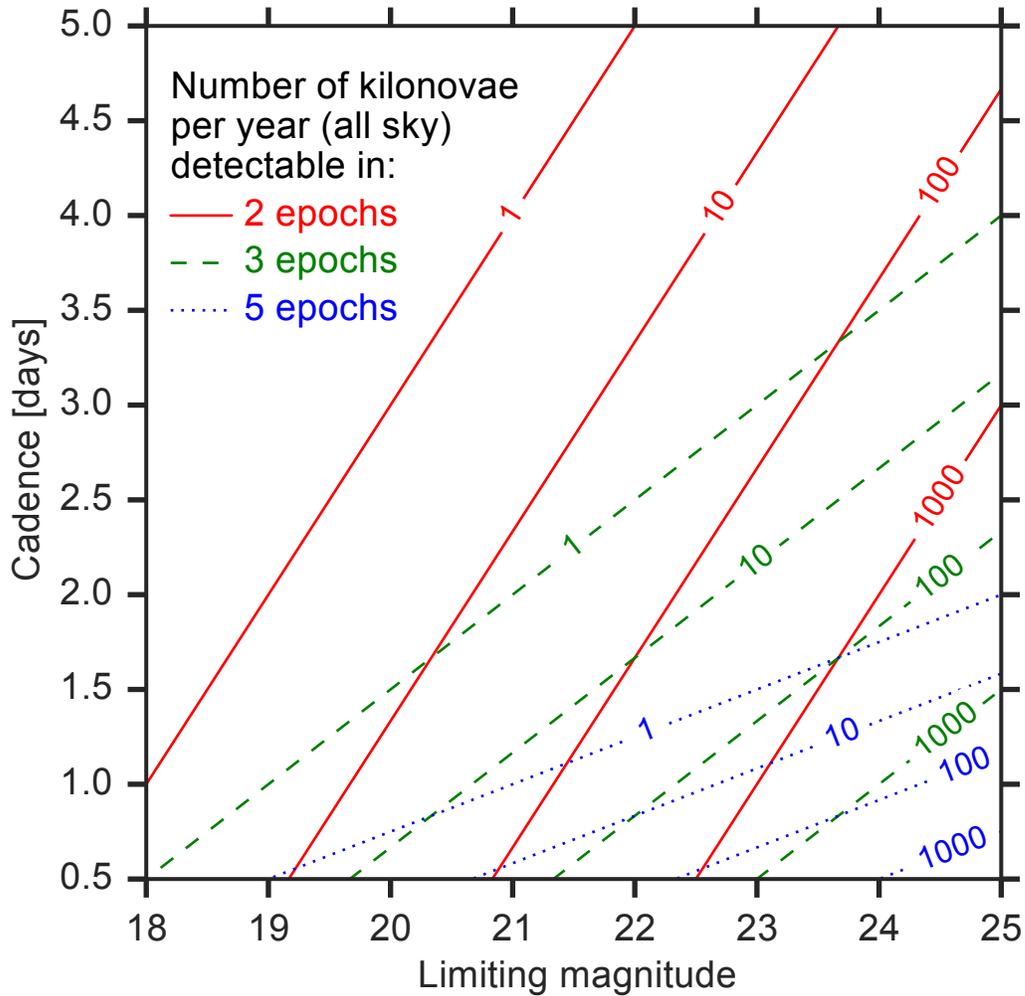

**Extended Data Figure 7 | Expected kilonova rates in optical transient surveys.** The number of AT 2017gfo-like events per year detectable by r-band transient surveys in two (solid lines), three (dashed lines) and five epochs (dotted lines) before fading from view. The numbers of events refer to the entire sky, and should be multiplied by the fraction of sky covered by the survey. We assume that the intrinsic rate of events is one per year out to 40 Mpc (scaling accordingly to larger distances).